
\centerline{\bf DISTORSION OF THE CMBR SPECTRUM:}

\centerline{\bf A TENTATIVE EXPLANATION}

\bigskip

\centerline{V. CELEBONOVIC$^1$, S. SAMUROVIC$^2$ and  M.M.
CIRKOVIC$^{3}$}
\bigskip

\centerline{\it $^1$ Institute of Physics, Pregrevica 118, 11080
Zemun--Beograd, Yugoslavia} 

\centerline{\it E-mail celebonovic@exp.phy.bg.ac.yu}

\bigskip
\centerline{\it $^2$ Public Observatory, Gornji Grad 16, Kalemegdan, 11000 
Beograd, Yugoslavia}

\centerline{\it E-mail srdjanss@afrodita.rcub.bg.ac.yu}
\bigskip

\centerline{\it $^3$ Dept. of Physics \& Astronomy; SUNY at Stony Brook,
Stony Brook, NY 11794-3800, USA}

\centerline{\it Astronomical Observatory, Volgina 7, 11000 Beograd, Yugoslavia}

\centerline{\it E-mail mcirkovic@aob.aob.bg.ac.yu}

\bigskip
\bigskip

\bigskip
\bigskip

\noindent{ABSTRACT: The purpose of this note is to present a possible explanation of the 
spectral distortions that could be measured in the Cosmic Microwave Background 
Radiation (CMBR) spectrum.  We propose that distorsions of the order of 
$y \sim 10^{-6}$ can be interpreted as a consequence of the radiative  decay of 
massive  tau neutrinos. We have also obtained a constraint on the value of the 
decaying neutrino mass.}

\bigskip
\bigskip

\noindent{INTRODUCTION}

\medskip

The CMBR represents perhaps the most important clue for 
understanding of  the universe as we know it today. It has a well established 
blackbody spectrum of temperature $T_0 = 2.728\pm 0.004$ (95 \% CL) 
(Fixsen {\it et al.} 1996). However, CMBR is not perfectly isotropic leading 
thus to very important conclusions concerning the evolution and development of 
the Universe. We here try to present the possible   
influence of decaying massive neutrinos  on the distorsions of the  CMBR 
profile.

 \bigskip

It is well known that the decay of stable particles,  for example, massive 
neutrinos decaying with a small branching ratio, distorts the CMBR spectrum 
(e.g., Smoot 1995). We assume that a  massive  neutrino $\nu _1$ of mass $m 
_{\nu _1}$ and lifetime $\tau$ decays into:

$$\nu _1\rightarrow \nu_2 + X \eqno(1)$$

or

$$\nu_1\rightarrow \gamma + \nu_2\eqno(2)$$

\noindent (e.g., Bernstein and Dodelson, 1990).

 The following relation holds:

$${m_{\nu_2}\over m_{\nu _1}}\ll 1.\eqno(3)$$

\bigskip
\noindent{CALCULATIONS}
\bigskip

In the decaying dark matter (DDM) theory (Sciama 1993, for modifications
see Sciama 1997) we assume that the mass of the $\tau$ neutrino is:

$$m_{\nu_\tau}=27.4\pm 0.2\, {\rm eV}.$$

As one can see this is a heavy constraint, leading thus to constraints
of some important cosmological parameters such as the density parameter
$\Omega$, Hubble constant $H_0$ and the age of the Universe (see Sciama 
1997)). One important characteristic of these decaying neutrinos is their
lifetime. According to some latest estimates it is:

$$\tau \sim 1\times 10^{23} \, {\rm s}\eqno(4)$$

\noindent (Sciama 1995, Sciama 1997).


\noindent{Fig. 1: {\it Dependence of $y$, deviation  of the CMBR spectrum from a
blackbody function, on the branching ratio $B$ and the mass of the tau
neutrino $m$.} }
\bigskip

Apart from being important for the structure of spiral galaxies (see e.g.,
Samurovic and Celebonovic 1996) these neutrinos could have
important influence on the ionization in the Universe (Sciama 1993). One of
the possible tests for their detection, besides the direct search for the
decay line originating from the emission of the line with energy $E_\gamma
\approx {m_{\nu _\tau}\over 2}\approx 13.7$ eV
(enough to ionize hydrogen)
that has recently started (mission EURD, April 21, 1997)\footnote {$^*$} {\tt
URL: http://www.laeff.esa.es/eng/laeff/activity/eurd.html.} could also be
 transfer of the decay photon energy to the CMBR photons
in a two step process: decay photons heat the electrons 
 through the inverse Compton scattering and hot electrons through inverse
 Compton scattering, give energy to the CMBR (Sunyaev-Zel'dovich effect). In 
the limit of small energy 
 transfer from decay photons to electrons, the deviations of the CMBR
 spectrum from a blackbody spectrum  $y\, \sim \, {\delta u\over 4u}\le 
 1.5\times 10^{-5}$ is:

$$y\simeq 6\times 10^{-8}B ({m_\nu\over  {\rm eV}})^2\eqno (5)$$

\noindent (Sethi 1997). As one can from the Figure 1 the possibly observable
region of $y$ ($y ^< _  \sim 4\times 10^{-6}$) requires that the branching ratio is
equal to 0.1. The mass of the tau neutrino $m _{\nu _\tau}=27.4\pm 0.2$ is
assumed. If these parameters $B$ and $m_ {\nu _\tau}$ are really in this range then the
evidence for the existence of these massive neutrinos will come in the next
decade when the new high-resolution CMBR experiments are planned. One can hope
that the evidence will come much earlier -- after the results of the EURD mission
are obtained.

According to Sethi (1997) (and references therein) the following equation 
holds:

$$y\simeq 5\times 10^{-5} B {m_\nu}^{4\over 3} t_\nu ^{-{1\over 
3}}.\eqno(6)$$

\noindent where $m _\nu$ is  expressed in eV and $t_\nu=\tau _\nu B$ is 
expressed in seconds.

Imposing $y=4\times 10^{-6}$ and $\tau _\nu \sim 10^{23}$ from Fig. 1 it 
follows from the equation (6) that: 

$$B\simeq {7.1\times 10^9\over m_\nu ^2}.\eqno(7)$$

Obviously $B$ must be $B\le 1$ thus giving from eq. (7):

$$m_\nu \le {8.4\times 10^4} {\rm eV}.\eqno(8)$$

A strange coincidence is that according to recent experiments $m _{\nu 
_\mu}\le 170$ keV (Brunner 1997)!
The current experimental upper limit (Brunner 1997) for the mass for the mass 
of the tau neutrino is:

$$ {\rm ALEPH:}\, \, \, m _{\nu _\tau} < 23.1 {\rm MeV}\, \, {\rm 95\% \; C.L.}$$

$$ {\rm OPAL:}\, \, \, m _{\nu _\tau} < 29.9 {\rm MeV}\, \, {\rm 95\% \; C.L.}$$

Taking the mean value and inserting it into eq. (7) one gets that 

$$B\simeq 10^{-5}.$$

\bigskip

\noindent{DDM AND THE REIONIZATION}
\bigskip

The only other astrophysical method in search for the DDM is direct observing of 
the decay line. Gunn-Peterson (GP) test suggests that this radiation is 
absorbed in ambiental intergalactic medium (IGM) or Lyman $\alpha$ forest 
clouds. Fluorescence will cause transfer of energy from original narrow line 
to optical recombination line of hydrogen. Unfortunately, contribution of DDM 
to Lyman $\alpha$ and H$\alpha$ lines is indistinguishable from fluorescence due to 
other photoionization sources. If  baryonic content of the Universe is near 
the lower bound from primordial nucleosynthesis (Walker {\it et al.} 1991), 
but much clumpier than usually supposed, internal photoionization sources 
(like star formation in galaxies) will dominate over the weak background. In 
any case, weakness of recombination signal from gas clouds free of internal 
ionization may be used to constrain lifetime and branching ratio of decaying 
neutrinos. Failure to detect any recombination emission at high and medium 
redshift  will probably be remedied soon, as sensitivity of deep space 
narrow-band imaging increases (Bland-Hawthorn 1997; Cirkovic and 
Samurovic 1997).

The exact picture of energy transport processes depends on (1) abundance of 
baryons in the Universe, and (2) clustering properties of IGM. The second 
issue remains basic problem for both optical and CMBR methods. In addition 
optical searches suffer from big uncertainty in value of metagalactic ionizing 
flux, $J(HI)$. "Classical" proximity effect value of $\sim 10^{-21}$ erg
s$^{-1}$ Hz$^{-1}$ cm$^{-2}$ sr$^{-1}$ (Bajtlik {\it et al.} 1988) is an 
overestimate, probably for a whole order of magnitude. This circumstance 
is another advantage of looking for DDM signature in the microwave 
anisotropies.

\bigskip
\noindent {CONCLUSION} 

\bigskip

We have shown that attempting to explain the distortions of the CMBR of the 
order of $\sim 10^{-6}$ as a consequence of the radiative decay of massive 
neutrinos   
is physically acceptable. Another conclusion is the upper limit for the 
mass of the neutrinos. The value we have obtained is in agreement with recent laboratory 
work and Sciama's theory. The continuation of this work aiming at predicting 
the values of observable parameters, such as angular scale and  correlation 
length, for a given set of cosmological parameters is in preparation.

\bigskip
\bigskip

\noindent {REFERENCES}

Bajtlik, S., Duncan, R.C. and Ostriker, J.P.: 1988, {\it Ap.J.}, {\bf 327}, 
570.

Bernstein, J. and Dodelson, S.: 1990, {\it Phys. Rev. D}, {\bf 41}, 354.

Bland-Hawthorn, J.: 1997, astro-ph 9704141.

Brunner, J.: 1997, preprint CERN-PPE/97-38.

Cirkovic, M.M. and Samurovic, S.: 1997, in preparation.

Fixsen, D.J.: 1996, {\it Ap.J.}, {\bf 473}, 576.

Samurovic, S. and Celebonovic, V.: 1996, {\it Publ. Astron. 
Belgrade}, {\bf 54}, 81.

Sciama, D.W.: 1993, {\it Modern Cosmology and the Dark Matter Problem}, Cambridge
University Press.

Sciama, D.W.: 1995, {\it Ap.J.}, {\bf 448}, 667.

Sciama, D.W.: 1997, astro-ph/9704081.

Smoot, G.F. {\it et al.}: 1995, astro-ph/9505139.

Sethi, S.K.:  1997, {\it Ap.J.}, {\bf 474}, 13.

Walker, T.P, Steigman, G., Schramm, D.N., Olive, K.N. and Kang, H.S.: 1991, 
{\it Ap.J.}, {\bf 376}, 51.

\bye